\documentclass[12pt,a4paper]{article}
\usepackage{epsfig}
\usepackage{graphicx}
\title{The study of the charged top-pion decay processes}

\author{Xuelei Wang , Wenna Xu, Linlin Du\\
 {\small College of
Physics and Information Engineering,}\\
\small{Henan Normal
University, Xinxiang  453002. P.R.China}
\thanks{This work is supported by the National Natural Science
Foundation of China(10175017), the Excellent Youth Foundation of
Henan Scientific Committee(02120000300), the Henan Innovation
Project for University Prominent Research Talents(2002kycx009).}
\thanks{E-mail:wangxuelei@263.net}
 }
\begin{document}
\maketitle

\begin{abstract}
\hspace{5mm} In the framework of top-color assisted
technicolor(TC2) theory, we study the four decay processes of
charged top-pion, i.e.,
 $\Pi^{+}_{t}\rightarrow t\overline{b}$,
$\Pi^{+}_{t}\rightarrow c\overline{b}$, $\Pi^{+}_{t}\rightarrow
W^{+}\gamma$, $\Pi^{+}_{t}\rightarrow W^{+}Z^{0}$, the decay
branching ratio of these modes are calculated. The results show
that the main decay channels of charged top-pion are the tree
level modes: $\Pi_t^+ \rightarrow t\bar{b}$ and $\Pi_t^+
\rightarrow c\bar{b}$. Light $\Pi_t^+$ is easier to be detected
than heavy one at future coliders. So, the study provides us some
useful information to search for charged top-pion.

\end{abstract}

\vspace{1.0cm} \noindent {\bf PACS number(s)}: 12.60Nz, 14.80.Mz,
12.15.LK, 14.65.Ha

\newpage
\noindent{\bf I. Introduction}~~\\

The large top quark mass suggests new dynamics, potentially
associated with electroweak symmetry breaking(EWSB). The
Glashow-Weinberg-Salam (GWS) theory has obtained great success in
describing the electromagnetic and weak interactions, but the
mechanism of electroweak symmetry breaking  is still unclear. So,
probing the mechanism of EWSB will be one of the most important
task both in the theoretical research and at the future high
energy colliders. To find the source of mass, we must find the
 cause of EWSB and the cause of flavor symmetry breaking. The
 standard Higgs Model, which bases on the gauge group
 $SU(3)_{C}\bigotimes SU(2)_{W}\bigotimes U(1)_{Y}$, accommodates the
 symmetry breakings, but the standard model doesn't explain the
 dynamic mechanism of the mass generation, and the scalar
 sector suffers from two serious problems: the gauge hierarchy
 and the triviality.

A new strong dynamics theory, technicolor(TC) model introduced by
Weinberg and Susskind\cite{TC}, offers a new insight into possible
 mechanisms of electroweak symmetry breaking. This kind of strong
 dynamical models of EWSB have evolved in the past dozens of
 years. In order to generate the masses of quarks and leptons, the extended technicolor has been introduced.
The simplest QCD-like extended technicolor model\cite{QCD-like}
leads to a too large oblique correction S
parameter\cite{S-parameter}, and is already ruled out by the CERN
$e^{+}e^{-}$ collider LEP precision electroweak measument data
\cite{LEP1,LEP2}. Various improvements have been proposed to make
the predictions consistent with the LEP precision measurement
data. As late as 1990s, we arrived at a viable model in which a
 new dynamic, topcolor,  can coexist and the top quark acquirs a
 dynamical mass through topcolor. Such model, called topcolor assisted
 technicolor model(TC2)\cite{TC2}, is consistent with the experimental limits
 and predicts a rich phenomenology that may be accessible to the colliders.

In TC2 theory, the electroweak symmetry breaking(ESB) is driven
mainly by technicolor interactions, the extended technicolor give
contributions to all ordinary quark and lepton masses including a
very small portion of the top quark mass: $m^{'}_{t}=\varepsilon
m_{t}$ $(0.03\leq \varepsilon \leq 0.1)$\cite{Burdman}. The
topcolor interactions also make small contributions to the ESB and
give rise to the main part of the top mass $(1-\varepsilon)m_{t}$
. One of the most general predictions of TC2 model is the
existence of three Pseudo-Goldstone Boson in a few hundred GeV
region, so called top-pions: $\Pi^{0}_{t}$,$\Pi^{\pm}_{t}$. The
physical particle top-pions can be regarded as the typical feature
of TC2 model. Thus, probing the possible signature of top-pions at
the high energy colliders is a good method to test TC2 model. In
order to determine which channel is the best one to to search for
top-pions, we need to know its decay branching ratio of each decay
modes.  The decay branching ratio of neutral top-pion has been
calculated  in Ref.\cite{yue}. The results show that the
flavor-changing mode $\Pi_t^0\rightarrow t\bar{c}$ is the best one
to search for neutral top-pion. In this paper, we calculate the
decay branching ratio of charged top-pion systematically which can
help us to search for charged top-pion.

This paper is organized as follow: In section II, we present the
calculation of the $\Pi^{+}_{t}$'s decay processes. The numerical
results of the decay processes and the conclusions are presented
in section III.

\noindent{\bf II. The calculation of the decay process of the $\Pi^{+}_{t}$ }~~\\

As it is known, for TC2 model, the underlying interaction,
topcolor  is non-universal, topcolor is preferrential to the third
generation particles. The couplings of top-pions to the fermions
are proportional to the masses of fermions, so, there exists a
large coupling $\Pi_t^+tb$. Furthermore, the flavor-changing
coupling $\Pi_t^+bc$ can be induced when one writes the
interactions in the quark mass eigen-basis.

The relative couplings of $\Pi^{+}_{t}$ can be written as
follow\cite{He}
 \begin{eqnarray}
  \frac{m_{t}}{\upsilon _{w}}\tan\beta [\sqrt{2}K_{UR}^{tt}K_{UL}^{bb*}
  \overline{t}_{R}b_{L}\Pi_{t}^{+} +\sqrt{2}K_{UR}^{tc}K_{UL}^{bb*}
  \overline{c}_{R}b_{L}\Pi_{t}^{+}]
\end{eqnarray}
where $\tan\beta=\sqrt{(\frac{\upsilon_{w}}{\upsilon_{t}})^2-1}$,
\hspace{0.5cm}$\upsilon_{t}\approx 60-100$ GeV is the top-pion
decay constant, $\upsilon_{w}=246$ GeV is the electroweak symmetry
breaking scale, $K^{bb}_{UL}$ is the matrix element of the unitary
matrix, $K_{UL}$ from which the Cabibbo-Kobayashi-Maskawa(CKM)
matrix can be derived as $V=K^{-1}_{UL}K_{UL}$, and $K^{ij}_{UR}$
are the matrix elements of the right-handed rotation matrix
$K_{UR}$, their values can be written as :
 \begin{eqnarray*}
  \hspace*{1cm}K^{bb*}_{UL}\approx 1 \hspace{1.5cm}K^{tt}_{UR}=1-\varepsilon
  \hspace*{1.5cm}K^{tc}_{UR}=\sqrt{2\varepsilon-\varepsilon^{2}}
  \end{eqnarray*}
  With above coupling, the charged top-pion can decay to $t\bar{b}$ and $c\bar{b}$ at tree level.
On the other hand, the charged top-pion can couple to a pair of
  gauge bosons through a triangle loop or a self-energy loop.
  Calculating the relative loops, we obtain the effective couplings of
  $\Pi^{+}_{t}W^{+}\gamma$ and $\Pi^{+}_{t}W^{+}Z^{0}$ as following:\\
\begin{eqnarray*}
 && \Pi^{+}_{t}W^{+}\gamma:\\
&&\frac{m_{t}^{2}\alpha_{e}}{6\pi\upsilon_{w}\sin\theta_{W}}\tan\beta(1-\varepsilon)
\{l_{\mu}p_{\nu}[2(2C_{22}-2C_{23}+C_{11}-C_{12}+C_{0})-(2C'_{22}-2C'_{23}
-C'_{11}\\
&&+C'_{12})] +g_{\mu\nu}[-2l\cdot
p(C_{11}-C_{12}+C_{0})+4C_{24}-l\cdot p(C'_{11}-C'_{12})
 +B'_{0}-2C'_{24}-3(B'_{1}+B'_{0})]\\
&&+i\varepsilon_{\mu\nu\rho\sigma}l^{\rho}p^{\sigma}[2(C_{11}-C_{12}+C_{0})+(C'_{11}-C'_{12})]
 \}
\end{eqnarray*}
\begin{eqnarray*}
&&\Pi^{+}_{t}W^{+}Z^{0}:\\
&&\frac{m_{t}^{2}\alpha_{e}\tan\beta(1-\varepsilon)}{4\pi\upsilon_{w}\sin^{2}\theta_{W}\cos\theta_{W}}
\{l_{\mu}p_{\nu}[(2C_{22}-2C_{23}-C_{12})
-\frac{4}{3}\sin^{2}\theta_{W}(2C_{22}-2C_{23}-C_{12}+C_{11}+C_{0})\\
&&-(1-\frac{2}{3}\sin^{2}\theta_{W})(2C'_{22}-2C'_{23}+C'_{12}-C'_{11})]
+g_{\mu\nu}[-B_{0}-m^{2}_{t}C_{0}-p^{2} (C_{11}-C_{12})+l\cdot
pC_{12}\\
&&+2C_{24}-\frac{4}{3}\sin^{2}\theta_{W}(l^{2}C_{12}+l\cdot
p(C_{12}-C_{11}-C_{0})+2C_{24})
-(1-\frac{2}{3}\sin^{2}\theta_{W})\\
&&(-B'_{0}+l\cdot p(C'_{11}-C'_{12})-l^{2}C'_{12}+2C'_{24})
+2\sin^{2}\theta_{W}(B'_{1}+B'_{0})]\\
&&-i\varepsilon_{\mu\nu\rho\sigma}l^{\rho}p^{\sigma}[(1-\frac{4}{3}\sin^{2}\theta_{W})(C_{12}-C_{11}-C_{0})
-(1-\frac{2}{3}\sin^{2}\theta_{W})(C'_{11}-C'_{12})]  \}
\end{eqnarray*}
where\hspace*{3cm} $C_{ij}=C_{ij}(-p,k,m_{t},m_{b},m_{t})$\\
\hspace*{4cm}$C'_{ij}=C'_{ij}(p,-k,m_{b},m_{t},m_{b})$ \\
\hspace*{4cm}$B_{i}=B_{i}(k,m_{b},m_{t})$\\
\hspace*{4cm}$B'_{i}=B_{i}(-k,m_{t},m_{b})$ \\
there, $B_i,C_{ij}$ is the standard 2-point and 3-point  scalar
integral. 'k' denotes  the momentum of incoming $\Pi^{+}_{t}$, 'p'
and 'l' denote the momenta of outcoming $W^{+}$ and $Z^0$(or
$\gamma$). $\theta_{W}$ is the
Weinberg angle.\\
\hspace*{1cm} With above discussion, we know that the charged
top-pion can also decay to $W^+Z$ and $W^+\gamma$ at loop level.
The Feynman diagrams of the decay modes are shown in Fig. 1.

As we know, $\Pi_t^+$ is a Pseudo Goldstone Boson and its spin is
zero, the outcoming particles must keep the total spin invariable.
So, in the channels of $\Pi^{+}_{t}\rightarrow W^{+}\gamma$ and
$\Pi^{+}_{t}\rightarrow W^{+}Z^{0}$, the helicity of the two
outcoming particles must be $(++), (- -), (00)$(Due to the  mass
of photon is zero, the case of (00) dose not exist in the decay
mode $\Pi^{+}_{t}\rightarrow W^{+}\gamma$).

With above discussion, we can directly write the amplitudes of
these channels
\begin{eqnarray*}
M_{\Pi^{+}_{t}\rightarrow
t\overline{b}}=\sqrt{2}\frac{m_{t}}{\upsilon_{w}}\tan\beta(1-\varepsilon)\overline{u_{t}}Lv_{b},
\end{eqnarray*}
\begin{eqnarray*}
M_{\Pi^{+}_{t}\rightarrow
c\overline{b}}=\sqrt{2}\frac{m_{t}}{\upsilon_{w}}\tan\beta
\sqrt{2\varepsilon-\varepsilon^{2}}\overline{u_{c}}Lv_{b},
\end{eqnarray*}
here, $L=\frac{1-\gamma_{5}}{2}$.\\
\begin{eqnarray*}
M^{++}_{\Pi^{+}_{t}\rightarrow W^{+}\gamma}&=&
\frac{-m_{t}\alpha_{e}}{6\pi\upsilon_{w}\sin\theta_{W}}\tan\beta(1-\varepsilon)
\{[-2l\cdot p(C_{11}-C_{12}+C_{0})+4C_{24}-l\cdot
p(C'_{11}-C'_{12})\\
&\ &+B'_{0}-2C'_{24}-3(B'_{1}+B'_{0})]
+\frac{1}{2}(M_{\Pi}^{2}-M_{W}^{2})[2(C_{11}-C_{12}+C_{0})+(C'_{11}-C'_{12})]\}
\end{eqnarray*}
\begin{eqnarray*}
M^{--}_{\Pi^{+}_{t}\rightarrow W^{+}\gamma}&=&
\frac{-m_{t}\alpha_{e}}{6\pi\upsilon_{w}\sin\theta_{W}}\tan\beta(1-\varepsilon)
\{[-2l\cdot p(C_{11}-C_{12}+C_{0})+4C_{24}-l\cdot
p(C'_{11}-C'_{12})\\
&\ &+B'_{0}-2C'_{24}-3(B'_{1}+B'_{0})]
-\frac{1}{2}(M_{\Pi}^{2}-M_{W}^{2})[2(C_{11}-C_{12}+C_{0})+(C'_{11}-C'_{12})]\}
\end{eqnarray*}
$|M_{\Pi^{+}_{t}\rightarrow W^{+}\gamma}|^{2}=|M^{++}_{\Pi^{+}_{t}\rightarrow W^{+}\gamma}|^{2}
+|M^{--}_{\Pi^{+}_{t}\rightarrow W^{+}\gamma}|^{2}$\\
\begin{eqnarray*}
M^{++}_{\Pi^{+}_{t}\rightarrow W^{+}Z}&=&
\frac{-m_{t}\alpha_{e}\tan\beta(1-\varepsilon)}{4\pi\upsilon_{w}\sin^{2}\theta_{W}\cos\theta_{W}}
\{-B_{0}-m^{2}_{t}C_{0}-p^{2}(C_{11}-C_{12})+l\cdot
pC_{12}+2C_{24}\\
&\ & -\frac{4}{3}\sin^{2}\theta_{W}[l^{2}C_{12}+l\cdot
p(C_{12}-C_{11}-C_{0})+2C_{24}]-(1-\frac{2}{3}\sin^{2}\theta_{W})[-B'_{0}\\
&&+l\cdotp(C'_{11}-C'_{12})-l^{2}C'_{12}+2C'_{24}]
+2\sin^{2}\theta_{W}(B'_{1}+B'_{0})\\
&-&\frac{1}{2}\sqrt{(M_{W}^{2}-M_{Z}^{2}+M_{\Pi}^{2})^{2}-4M_{\Pi}^{2}M_{W}^{2}}
[(1-\frac{4}{3}\sin^{2}\theta_{W})(C_{12}\\
&\
&-C_{11}-C_{0})-(1-\frac{2}{3}\sin^{2}\theta_{W})(C'_{11}-C'_{12})]\}
\end{eqnarray*}
\begin{eqnarray*}
M^{--}_{\Pi^{+}_{t}\rightarrow W^{+}Z}&=&
\frac{-m_{t}\alpha_{e}\tan\beta(1-\varepsilon)}{4\pi\upsilon_{w}\sin^{2}\theta_{W}\cos\theta_{W}}
\{-B_{0}-m^{2}_{t}C_{0}-p^{2}(C_{11}-C_{12})+l\cdot
pC_{12}+2C_{24}\\
&\ & -\frac{4}{3}\sin^{2}\theta_{W}[l^{2}C_{12}+l\cdot
p(C_{12}-C_{11}-C_{0})+2C_{24}]-(1-\frac{2}{3}\sin^{2}\theta_{W})\\
&&[-B'_{0}+l\cdot p(C'_{11}-C'_{12})-l^{2}C'_{12}+2C'_{24}]
+2\sin^{2}\theta_{W}(B'_{1}+B'_{0})\\
&+&\frac{1}{2}\sqrt{(M_{W}^{2}-M_{Z}^{2}+M_{\Pi}^{2})^{2}-4M_{\Pi}^{2}M_{W}^{2}}
[(1-\frac{4}{3}\sin^{2}\theta_{W})\\
&\
&(C_{12}-C_{11}-C_{0})-(1-\frac{2}{3}\sin^{2}\theta_{W})(C'_{11}-C'_{12})]\}
\end{eqnarray*}
\begin{eqnarray*}
M^{00}_{\Pi^{+}_{t}\rightarrow W^{+}Z}&=&
\frac{-m_{t}\alpha_{e}\tan\beta(1-\varepsilon)}{4\pi\upsilon_{w}\sin^{2}\theta_{W}\cos\theta_{W}}
\{\frac{(M_{W}^{2}-M_{Z}^{2}+M_{\Pi}^{2})^{2}-4M_{\Pi}^{2}M_{W}^{2}}{4M_{W}M_{Z}}[2C_{22}-2C_{23}-C_{12}\\
&\
&-\frac{4}{3}\sin^{2}\theta_{W}(2C_{22}-2C_{23}-C_{12}+C_{11}+C_{0})
-(1-\frac{2}{3}\sin^{2}\theta_{W})
(2C'_{22}-2C'_{23}\\
&&+C'_{12}-C'_{11})]
+\frac{1}{4M_{W}M_{Z}M{\Pi}^{2}}[M_{\Pi}^{4}-(M_{W}^{2}-M_{Z}^{2})^{2}+(M_{W}^{2}-M_{Z}^{2}+M_{\Pi}^{2})^{2}\\
&\
&-4M_{\Pi}^{2}M_{W}^{2}][-B_{0}-m^{2}_{t}C_{0}-p^{2}(C_{11}-C_{12})+l\cdot
pC_{12}\\
&\ & +2C_{24}-\frac{4}{3}\sin^{2}\theta_{W}(l^{2}C_{12}+l\cdot
p(C_{12}-C_{11}-C_{0})\\
&\ &+2C_{24})-(1-\frac{2}{3}\sin^{2}\theta_{W})(-B'_{0}+l\cdot
p(C'_{11}-C'_{12})\\
&\ &-l^{2}C'_{12}+2C'_{24})+2\sin^{2}\theta_{W}(B'_{1}+B'_{0})]\}
\end{eqnarray*}
$|M_{\Pi^{+}_{t}\rightarrow W^{+}Z^{0}}|^{2}=|M^{++}_{\Pi^{+}_{t}\rightarrow W^{+}Z^{0}}|^{2}
+|M^{--}_{\Pi^{+}_{t}\rightarrow W^{+}Z^{0}}|^{2}
+|M^{00}_{\Pi^{+}_{t}\rightarrow W^{+}Z^{0}}|^{2}$\\

\noindent{\bf III. The numerical results and conclusions }~~ \\

To determine which channel is the ideal one to search for the
charged top-pion, we need to calculate the decay branching ratio
of each decay mode. The decay branching ratio can be defined as:
$BR=\Gamma_{branch}/\Gamma_{total}$. In our calculations, we take
$m_{t}=174$ GeV, $M_{Z}=91.187$ GeV, $m_{b}=4.9$ GeV, $m_{c}=1.5$
GeV, $M_{W}=80.4$ GeV, $\upsilon_{t}=60$ GeV, and
$\sin^{2}\theta_{W}=0.23$. The electromagenetic fine structure
constant $\alpha_{e}$ is taken as $\alpha_{e}=1/128.9$. There are
two free parameters in the expressions of the decay branching
ratio: $\varepsilon, M_{\Pi}$. To see the influence of these
parameters on the decay branching ratio, we take the mass of the
top-pion ($M_{\Pi}$) to vary from 200 GeV to 400 GeV and
$\varepsilon=0.03,0.06,0.1$, respectively. The numerical results
are summarized in Fig.2-6.

The Fig.2-4 are the plots of the decay branching ratio as the
function of $M_{\Pi}$ for $\varepsilon=0.03,0.06,0.1$,
respectively. The results show that the main decay channels are
the tree level modes:$\Pi_t^+ \rightarrow t\bar{b}$ and $\Pi_t^+
\rightarrow c\bar{b}$. With $M_{\Pi}$ increasing, the decay
branching ratio $Br(\Pi_t^+ \rightarrow t\bar{b})$ increases but
Br($\Pi_t^+ \rightarrow c\bar{b})$ decreases inversely. Although
the coupling $\Pi_t^+t\bar{b}$ is  much stronger than the coupling
$\Pi_t^+c\bar{b}$, the small $M_{\Pi}$ depresses the decay width
of $\Pi_t^+ \rightarrow t\bar{b}$ very much, so, the decay widths
of two processes are at the same level for small $M_{\Pi}$. The
background of $c\bar{b}$ channel should be more clean than that of
$t\bar{b}$ channel. Thus, for light $\Pi_t^+$, $c\bar{b}$ is the
best channel to detect $\Pi_t^+$. For heavy $\Pi_t^+$, $\Pi_t^+
\rightarrow t\bar{b}$ is the dominent decay mode, we should detect
charged top-pion via $t\bar{b}$ channel, but the background of
$t\bar{b}$ channel is large. On the other hand, the total decay
width is very large for heavy charged top-pion, so, it is more
difficult for us to detect heavy charged top-pion.

Comparing with the tree level process, the decay branching ratio
of loop level processes $\Pi_t^+ \rightarrow W^+Z^0(\gamma)$ are
very small. The polarized state of the outcoming particles can
effect the decay widths. The decay widths arising from different
polarized states are shown in Fig.5-6. From Fig.5, we can see that
the main contribution to the decay width arise from polarized
state (0,0). But in the decay mode $\Pi^{+}_{t}\rightarrow
W^{+}\gamma$, the polarized state (00) dose not exist, so, the
decay width of this decay mode is about two order of magnitude
smaller than that of $\Pi^{+}_{t}\rightarrow W^{+}Z^{0}$. The
decay width of $\Pi^{+}_{t}\rightarrow W^{+}Z^{0}$ can reach the
level of a few GeV.

 In conclusion, we study the decay processes of the charged
top-pion in the TC2 model systematically. The results show that
the main decay channels of charged top-pion are the tree level
modes: $\Pi_t^+ \rightarrow t\bar{b}$ and $\Pi_t^+ \rightarrow
c\bar{b}$. The decay widths of two processes are at the same level
for small $M_{\Pi}$. The decay branching ratio of loop level
processes $\Pi_t^+ \rightarrow W^+Z^0(\gamma)$ are very small. For
light $\Pi_t^+$, $c\bar{b}$ is the best channel to detect
$\Pi_t^+$. It is more difficult to detect heavy charged top-pion
with large background and large total decay width.

\newpage
\begin{figure}[h]
\begin{center}
\epsfig{file=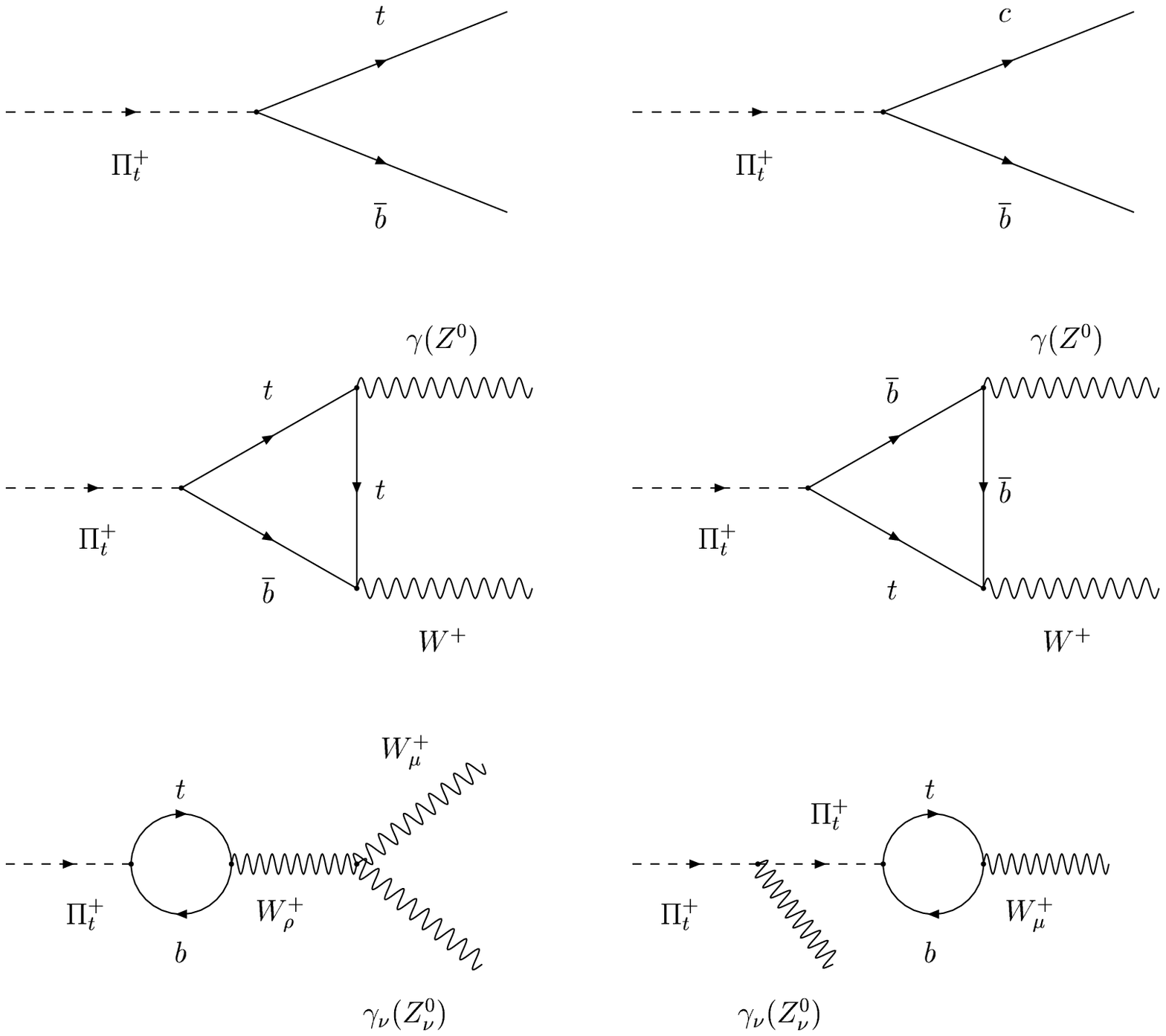,width=500pt,height=600pt} \vspace{-8.5 cm}
\caption{ The Feynman diagrams of the decay processes of
    $\Pi^{+}_{t}$.}
\label{fig1}
\end{center}
\end{figure}

\newpage
\begin{figure}[h]
\begin{center}
\includegraphics [scale=1.3] {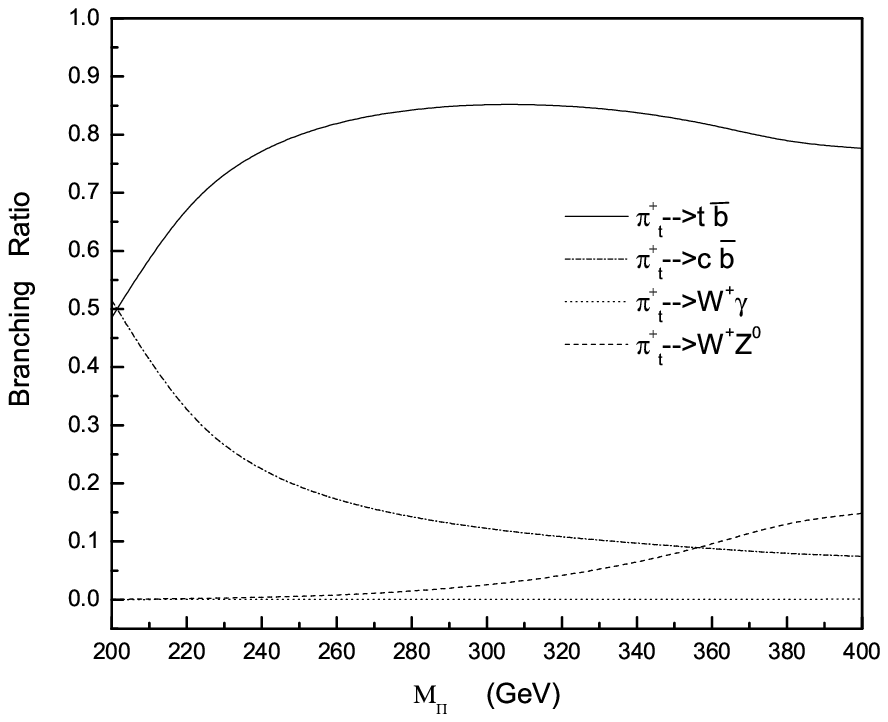}
\caption{ The decay branching ratio of $\Pi_t^+$ for
$\varepsilon=0.03$.} \label{fig2}
\end{center}
\end{figure}

\begin{figure}[h]
\begin{center}
\includegraphics [scale=1.3] {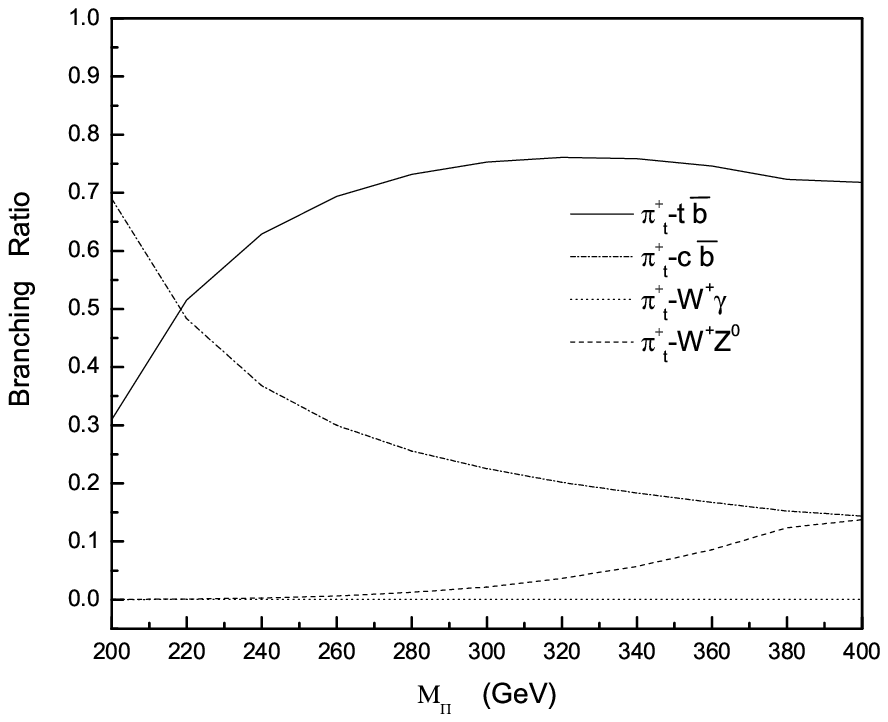}
\caption{ The decay branching ratio of $\Pi_t^+$ for
$\varepsilon=0.06$.} \label{fig.3}
\end{center}
\end{figure}

\newpage
\begin{figure}[h]
\begin{center}
\includegraphics [scale=1.3] {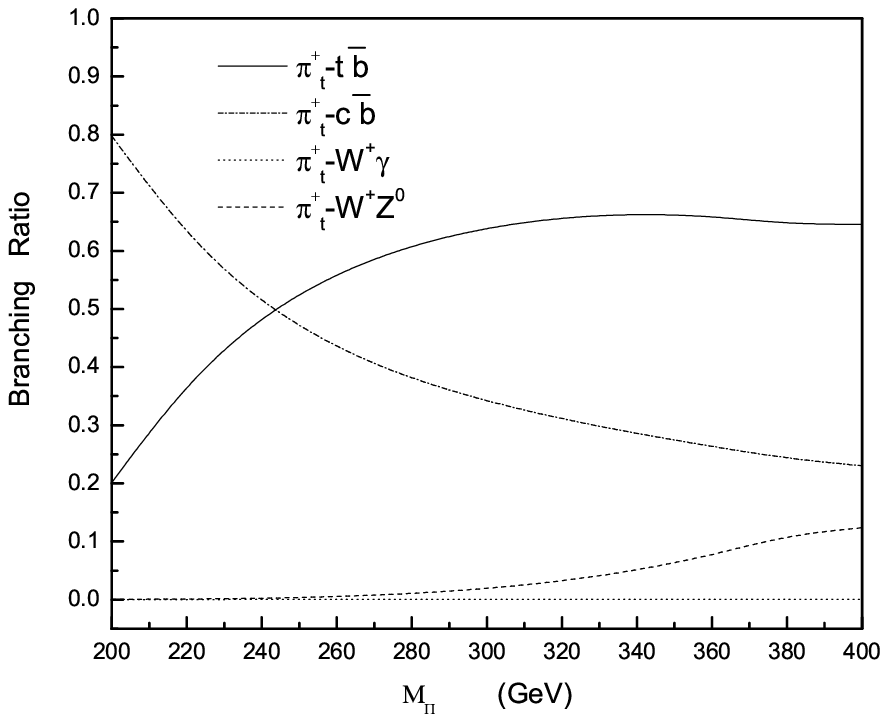}
\caption{The decay branching ratio of $\Pi_t^+$ for
$\varepsilon=0.1$.} \label{fig.4}
\end{center}
\end{figure}

\newpage
\begin{figure}[h]
\begin{center}
\includegraphics [scale=1.3] {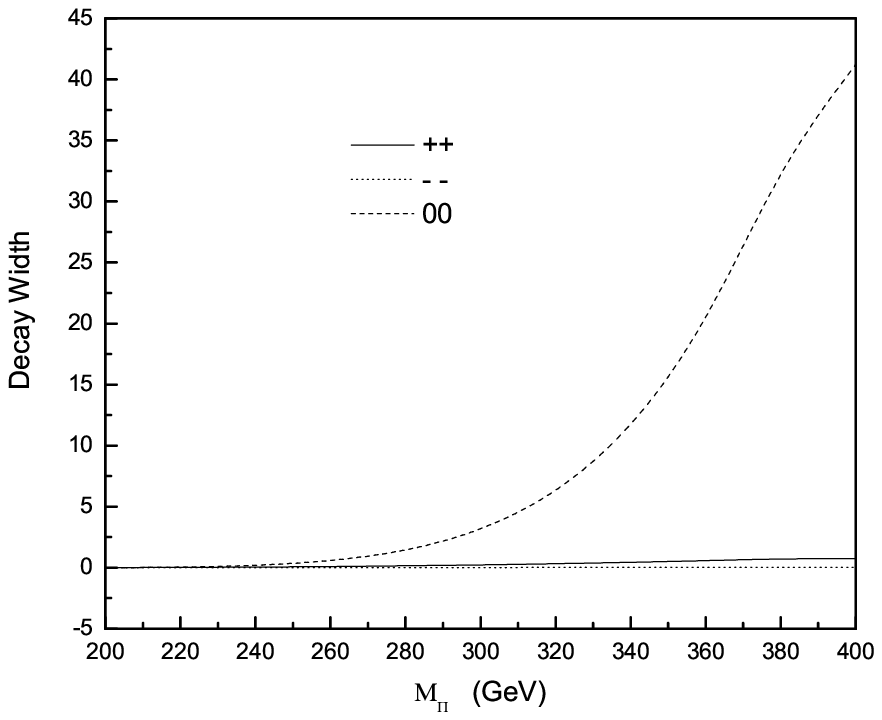}
\caption{ The decay width of $\Pi_t^+\rightarrow W^+Z$ varying as
the helicity of the outcoming particles($\varepsilon=0.06$). }
\label{fig.5}
\end{center}
\end{figure}
\newpage

\begin{figure}[h]
\begin{center}
\includegraphics [scale=1.3] {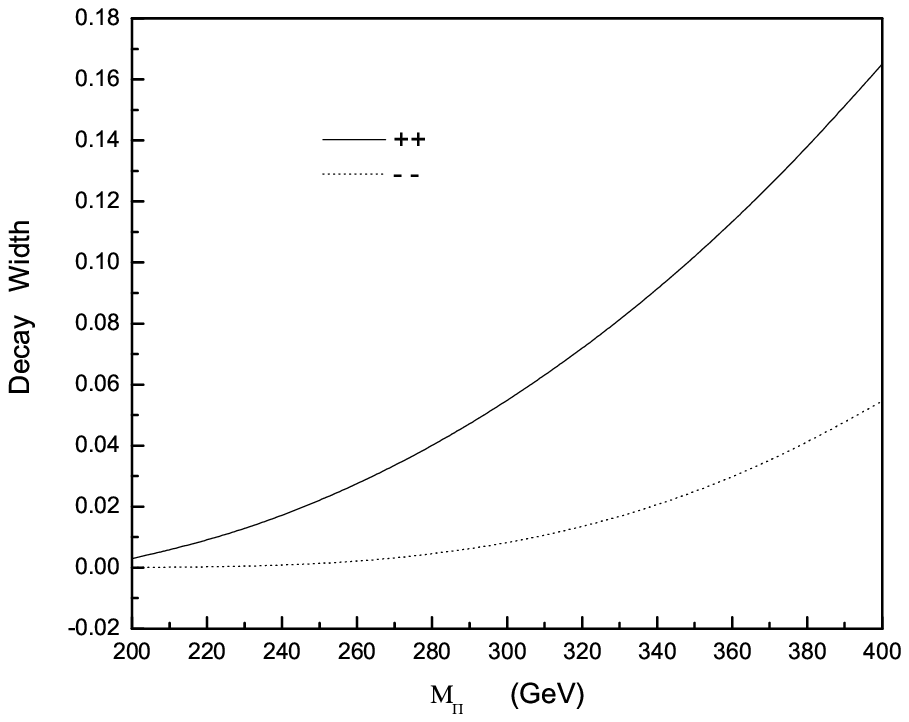}
\caption{ The decay width of $\Pi_t^+\rightarrow W^+\gamma$
varying as the helicity of the outcoming
particles($\varepsilon=0.06$). } \label{fig.6}
\end{center}
\end{figure}


\newpage
\begin{thebibliography}{100}
\bibitem{TC}
 S. Weinberg, {\it Phys. Rev.} D{\bf 13}, (1976)974;
{\bf 19}, (1979)1277; L.Susskind, {\em Phys. Rev. D}{\bf
20},2619(1979).
\bibitem{QCD-like}
S. Dimopoulos and L. Susskind, {\it Nucl.Phys.} B{\bf 155};
237(1979); E. Eichten and K. Lane, {\it Phys. Lett.} {\bf 90}B,
125(1980).
\bibitem{S-parameter} M. Peskin and T. Takeuchi, {\it Phys. Rev. Lett.}
{\bf 65}, 964(1990).
\bibitem{LEP1}
 J. Erler and P. Langacker in Review of Particle
Physics, {\it Eur. Phys.} J. C{\bf 3}, 90(1998).
\bibitem{LEP2}
K. Hagiwara, D. Haidt and S. Matsumoto, {\it Eur. Phys.} J. C{\bf
2},
 95(1995).
\bibitem{TC2}
 C. T. Hill, {\it Phys. Lett.} B{\bf 345}, (1995)483;
K. Lane and E. Eichten, {\it Phys. Lett.} B{\bf 352}, (1995)382;
K. Lane, {\it Phys. Rev.} D{\bf 54}, (1996)2204; R. S. Chivukula,
B. A. Dobrescu, H. Georgi and C. T. Hill, {\it Phys. Rev.} D{\bf
59}, (1999)075003.
\bibitem{Burdman}
G. Buchalla, G. Burdman, C. T. Hill, D. Kominis, {\it Phys. Rev.}
D{\bf 53}, 5185(1996).
\bibitem{yue}
C.X.Yue, Q.J.Xu, G.L.Liu, J.T.Li,{\it Phys. Rev. D}{\bf
63},(2001)115002.
\bibitem{He}
Hong-Jian He, Shinya Kanemura and C. P. Yuan, hep-ph/0203090.
\end{thebibliography}
\end{document}